\documentclass[11pt]{article}

\usepackage{latexsym,amssymb,amsfonts,graphicx,makeidx,rotating,ifthen,amsmath,subfigure,epsfig,slashbox,a4wide}

\newcommand{\MSOB}{\text{MSO}_2}
\newcommand{\X}{\times}
\newcommand{\val} {\mbox{val}}

\newcommand{\TW} {\mbox{TW}}
\newcommand{\NLC} {\mbox{NLC}}

\newcommand{\MSOA}{\mbox{MSO}_1}
\newcommand{\nlcw} {\mbox{NLC-width}}

\newcommand{\lab} {\mbox{lab}}

\newcommand{\CW} {\mbox{CW}}
\newcommand{\cw} {\mbox{clique-width}}

\newcommand{\tw} {\mbox{tree-width}}
\newcommand{\graph} {\text{val}}

\newcommand{\col} {\text{col}}

\newtheorem{theorem}{Theorem}[section]
\newtheorem{example}[theorem]{Example}

\newtheorem{definition}[theorem]{Definition}

\newtheorem{problem}[theorem]{Problem}

\begin{document}
\title{A comparison of two approaches for polynomial time algorithms computing basic graph parameters\thanks{Small parts of this paper 
have been published in an extended abstract \cite{EGW01a}.} \thanks{This paper is a summary of two chapters of \cite{Gur07e}.}}

\author{Frank Gurski\thanks{Heinrich-Heine Universit\"at D\"usseldorf,
Department of Computer Science, D-40225 D\"usseldorf, Germany,
E-Mail: gurski-corr@acs.uni-duesseldorf.de, }}

\maketitle

\begin{abstract}
In this paper we compare and illustrate the algorithmic use of graphs of bounded tree-width
and graphs of bounded clique-width. For this purpose we  give polynomial time algorithms for computing the four basic graph parameters independence number, clique number,
chromatic number, and clique covering number on a given tree structure of  graphs of bounded tree-width
and graphs of bounded clique-width
in polynomial time. We also present linear time algorithms for computing the latter four basic graph parameters on trees, i.e. graphs of tree-width 1, and on co-graphs, i.e. graphs of clique-width at most 2. 

\end{abstract}


\section{Introduction\label{sec-intro}}

A graph parameter is a mapping that associates every graph with a positive integer. Well known graph parameters are independence number, dominating number, and chromatic number. 
In general the computation of such parameters for some given graph is NP-hard. In this work we give
fixed-parameter tractable (fpt) algorithms for computing basic graph parameters
restricted to graph classes of bounded tree-width and graph classes of bounded clique-width.

The tree-width of graphs has been defined in 1976 by Halin  \cite{Hal76}
and independently in 1986 by Robertson and Seymour  \cite{RS86} by the existence of a tree decom\-po\-sition. Intuitively, the tree-width of some graph $G$ measures how far $G$ differs from a tree.

Two more powerful and more recent graph parameters are clique-width\footnote{The operations in the definition of
the graph parameter clique-width were first considered by Courcelle, Engelfriet, and Rozenberg in \cite{CER91} and \cite{CER93}.} and NLC-width\footnote{The abbreviation NLC results from the 
node label controlled embedding mechanism originally defined for graph
grammars \cite{ER97}.} both
defined in 1994, by Courcelle and Olariu  \cite{CO00} and by Wanke  \cite{Wan94}, respectively.
The clique-width of a graph $G$ is the least integer $k$ such that 
$G$ can be defined by operations on vertex-labeled graphs using $k$ labels. 
These operations are the vertex disjoint
union, the addition of edges between vertices controlled by a label pair, and
the relabeling of vertices. The NLC-width of a graph $G$ is defined similarly in terms
of closely related operations.
The only essential difference between the composition mechanisms of clique-width
bounded graphs and NLC-width bounded
graphs is the addition of edges. In an NLC-width composition the addition of edges
is combined with the union operation.  
Intuitively, the clique-width and NLC-width of some graph $G$ measure how far $G$ or its edge
complement graph differs from a clique (i.e. a complete graph).

See \cite{BK07} and \cite{HOSG07} for two recent surveys on tree-width and clique-width.

One of the main reasons for regarding tree-width and clique-width is that a lot of hard problems
become solvable in polynomial when restricted to graph classes of bounded tree-width and graph classes of bounded clique-width.

In this paper we present two dynamic programming schemes to solve graph problems on a given tree decomposition 
(Chapter \ref{ch9}) and graph problems on a given clique-width expression (Chapter \ref{ch10}).
These and similar dynamic programming approaches have been used in \cite{Arn85},\cite{AP89},\cite{Bod87},\cite{Bod88}, \cite{Bod90},\cite{Hag00},\cite{KZN00},\cite{ZFN00},\cite{INZ03} to solve a large number of NP-complete graph problems on graph classes of bounded tree-width
and in \cite{Wan94}, \cite{EGW01a}, \cite{GK03},\cite{KR03},\cite{Tod03},\cite{GW06}, \cite{MRAG06},\cite{Rao06},\cite{ST07},\cite{Rao07a}, \cite{Gur07e} to solve a large number of NP-complete graph problems on graph classes of bounded clique-width.
We apply our two approaches in order to compute the four basic graph parameters independence number, clique number,
chromatic number, and clique covering number on a given tree structure
in polynomial time. It is well known that the computation of all four parameters is NP-complete on general
graphs \cite{GJ79}. The running time of our algorithms is exponential in the tree-width or clique-width $k$ but 
polynomial in the instance size. 
Thus if we restrict our problems to graph classes of bounded widths, parameter $k$ will occur as a constant in the running time and we obtain polynomial time parameterized complexity algorithms, see the books \cite{Nie06}, \cite{FG06}, and \cite{DF99} for surveys. 
We also present linear time algorithms for computing the latter four basic graph parameters on trees, i.e. graphs of tree-width 1, and on co-graphs, i.e. graphs of clique-width at most 2. 

Regarding theoretically results from monadic second order logic \cite{CMR00,CM93},
the existence of the solutions for computing the independence number
and  clique number on graphs of bounded tree-width or graphs of bounded clique-width is known. Nevertheless our shown dynamic programming solutions on a given tree structure are more feasible. The same remark holds true regarding complement problems on graphs of bounded clique-width. Further, this paper compares the two main approaches which are
used to solve graph problems on tree-structured graph classes.

Finally, in Section \ref{ch-concl} we discuss the  vertex cover number and the dominating number as two further well known graph parameters which can be computed in polynomial  time on graphs of bounded tree-width and graphs of bounded clique-width.  Further we
stress that both given dynamic programming approaches to  solve problems along a tree decomposition
and along a clique-width expression are useful.


\section{Preliminaries}


\subsection{Definitions of graph parameters with algorithmic applications}

One of the most famous tree structured graph classes are graphs of bounded
tree-width. The notion of tree-width was defined in the 1980s by Robertson and 
Seymour in \cite{RS86} as follows.

\begin{definition}[$\TW_k$, tree-width, \cite{RS86}]\label{D0}
A {\em tree de\-com\-position} of a graph $G=(V_G,E_G)$ is a pair $({\mathcal X},T)$
where $T=(V_T,E_T)$ is a tree and ${\mathcal X}=\{X_{u} \mid u\in V_T\}$ is a
family of subsets $X_u \subseteq V_G$, one for each node $u$ of $T$, such that
the following three conditions hold true.
\begin{itemize}
\item
$\cup_{u \in V_T} X_u$ = $V_G$.
\item
For every edge $\{v_1,v_2\}\in E_G$, there is some node $u \in V_T$ such that
$v_1 \in  X_u$ and $v_2 \in X_u$.
\item
For every vertex $v \in V_G$ the subgraph of $T$ induced by the nodes $u \in
V_T$ with $v \in X_u$ is connected.
\end{itemize}
The {\em width} of a tree decomposition $({\mathcal X}=\{X_u \mid u\in V_T\},T=(V_T,E_T))$
is $\max_{u \in V_T} |X_u|-1$. The {\em tree-width} of a graph $G$ (denoted by $\tw(G)$) is the
smallest integer $k$ such that there is a tree decomposition 
$({\mathcal X},T)$ for $G$ of width $k$.
By $\TW_k$ we denote the set of all graphs of tree-width at most $k$.
\end{definition}

\begin{figure}[ht]
\centerline{\epsfig{figure=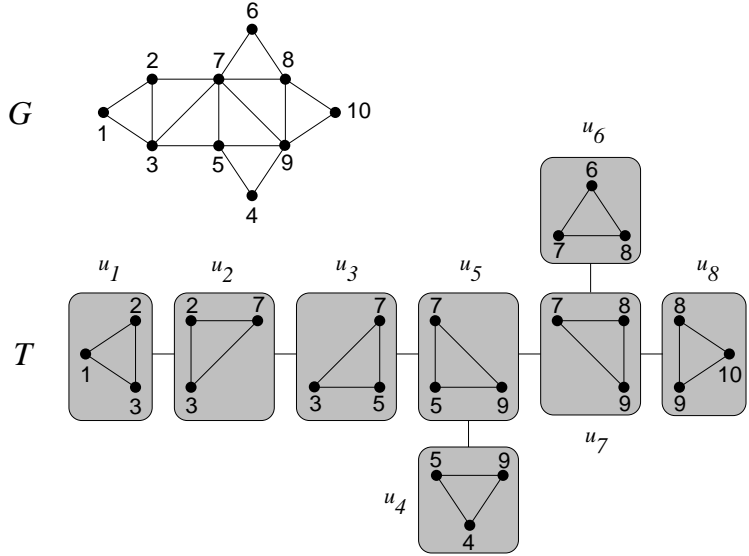,width=280pt}}
\caption[A graph and a corresponding  tree decomposition] {\label{F1} A graph $G$ of tree-width 2 and a tree decomposition $({\mathcal X},T)$ of width $2$ for $G$.}
\end{figure}

Fig. \ref{F1} shows a graph $G$ and a tree decomposition of width 2 for $G$.

Next we give some examples for graph classes of bounded tree-width.
Trees have tree-width 1 \cite{Bod98}.  Series parallel graphs have tree-width at most 2 \cite{WC83}.  Halin graphs have tree-width at most 3 \cite{Bod88c}. $k$-outerplanar graphs have tree-width at most $3k-1$ \cite{Bod88c}.
A more detailed overview on graph classes of bounded tree-width can be found in \cite{Bod86}, \cite{Bod88c}. 

On the other hand, the tree-width of complete graphs and thus of co-graphs (which are defined in Section \ref{sec-co}) is
not bounded \cite{Bod98}. 

\bigskip
Two more recent parameters are clique-width and NLC-width, which are originally defined
for labeled graphs.

Let $[k]:=\{1,\ldots,k\}$ be the set of all integers between $1$ and $k$.
We work with finite undirected labeled {\em graphs} $G=(V_G,E_G,\lab_G)$,
where $V_G$ is a finite set of {\em vertices} labeled by some mapping
$\lab_G: V_G \to [k]$ and $E_G \subseteq \{ \{u,v\} \mid u,v \in
V_G, u \not= v\}$ is a finite set of {\em edges}.
A labeled graph  $J=(V_J,E_J,\lab_J)$ is a {\em subgraph} of $G$ if $V_J
\subseteq V_G$, $E_J \subseteq E_G$ and $\lab_J(u)=\lab_G(u)$ for all $u \in
V_J$. $J$ is an {\em induced subgraph} of $G$ if additionally $E_J=\{ \{u,v\} \in E_G
\mid u,v \in V_J\}$. The labeled graph consisting of a single vertex labeled by $a \in [k]$ is denoted by $\bullet_a$.

The notion of clique-width of labeled graphs is defined by Courcelle and
Olariu in \cite{CO00} as follows.

\begin{definition}[$\CW_k$, $\cw$, \cite{CO00}]
\label{D1a}
Let $k$ be some positive integer. The class $\CW_k$ of labeled graphs is
recursively defined as follows.
\begin{itemize}
\item
The single vertex $\bullet_a$ labeled by some $a \in [k]$ is in $\CW_k$.
\item
Let $G \in \CW_k$ and $J \in \CW_k$ be two
vertex disjoint  labeled graphs. Then $G \oplus J:=(V',E',\lab')$
defined by $V':=V_G  \cup V_J$, $E':=E_G \cup E_J$, and
\[\lab'(u) \ := \ \left\{\begin{array}{ll}
\lab_G(u) & \mbox{if } u \in V_G\\
\lab_J(u) & \mbox{if } u \in V_J\\
\end{array} \right.,\ {\rm{for~all~}}u\in V'    \]
is in $\CW_k$.
\item
Let $a,b \in [k]$ be two distinct integers and $G \in
\CW_k$ be a labeled graph then
\begin{itemize}
\item $\rho_{a \rightarrow b}(G):=(V_G,E_G,\lab')$ defined by
\[\lab'(u) \ := \ \left\{\begin{array}{ll}
\lab_G(u) & \mbox{if } \lab_G(u) \not= a\\
        b & \mbox{if } \lab_G(u) = a\\
\end{array} \right.,\ {\rm{for~all~}}u\in V_G\]
is in $\CW_k$ and
\item
$\eta_{a,b}(G)\ :=\ (V_G,E',\lab_G)$ defined by
\[E':=E_G \cup \{ \{u,v\} \mid u,v \in V_G,~u\not=v,~\lab(u)=a,~\lab(v)=b \}\]
is in $\CW_k$.
\end{itemize}
\end{itemize}
The  {\em $\cw$} of a labeled graph $G$ (denoted by $\cw(G)$)  is the least integer $k$ such that
$G \in \CW_k$.
\end{definition}

The notion of NLC-width of labeled graphs is defined by Wanke in \cite{Wan94} as follows.

\begin{definition}[$\NLC_k$, $\nlcw$, \cite{Wan94}]
\label{D1}Let $k$ be some positive integer. 
The graph class $\NLC_k$ of labeled graphs is
recursively defined as follows.
\begin{enumerate}
\item
The single vertex graph $\bullet_a$ for some $a \in [k]$ is in $\NLC_k$.
\item
Let $G=(V_G,E_G,\lab_G) \in \NLC_k$ and $J=(V_J,E_J,\lab_J) \in \NLC_k$ be
two vertex disjoint  labeled graphs and $S \subseteq [k]^2$ be a relation, then
$G \times_S J := (V',E',\lab')$ defined by $V':=V_G \cup V_J$,
\[E':=E_G \cup E_J \cup \{\{u,v\} \mid u \in V_G,~v \in
V_J,~(\lab_G(u),\lab_J(v)) \in S \},\] and
\[\lab'(u) \ := \
\left\{\begin{array}{ll} \lab_G(u) & \mbox{if } u \in V_G\\
\lab_J(u) & \mbox{if } u \in V_J\\
\end{array} \right.,~\forall u \in V'\]
is in $\NLC_k$.
\item
Let $G=(V_G,E_G,\lab_G) \in \NLC_k$ be a labeled graph and $R:[k] \to [k]$ be a function, then
$\circ_R(G) := (V_G,E_G,$ $\lab')$ defined by
$\lab'(u) := R(\lab_G(u)),~\forall u \in V_G$ is in $\NLC_k$.
\end{enumerate}
The {\em NLC-width} 
of a  labeled graph $G$  (denoted by $\nlcw(G)$) is the least integer $k$ such that
$G \in \NLC_k$.
\end{definition}

An expression built with the operations
$\bullet_a,\oplus,\rho_{a \rightarrow b},\eta_{a,b}$ for integers $a,b \in
[k]$ is called a {\em clique-width $k$-expression}.
An expression $X$ built with the operations $\bullet_a,\X_S,\circ_R$ for $a \in [k]$, $S \subseteq [k]^2$, and $R:[k] \to [k]$   is called an {\em NLC-width $k$-expression}.
The {\em clique-width} (the {\em NLC-width}) of an unlabeled graph $G=(V,E)$ is the smallest integer
$k$, such that there is some mapping  $\lab : V \to [k]$ such that
the labeled graph  $(V,E,\lab)$  has clique-width at most $k$ (NLC-width at most $k$, respectively). 
The graph defined by expression $X$ is denoted by $\val(X)$. 
By the definition of $k$-expressions it is easy to verify that
graphs of bounded clique-width and graphs of bounded NLC-width are closed under taking induced subgraphs.

Every clique-width $k$-expression $X$ has by its recursive definition a tree structure that is
called the {\em clique-width $k$-expression-tree} $T$ for $X$.
$T$  is an ordered rooted tree whose
leaves correspond to the vertices of graph $\val(X)$ and the inner nodes correspond to
the operations of $X$, see \cite{EGW03}.
In the same way we  define the NLC-width $k$-expression-tree for every NLC-width 
$k$-expression, see \cite{GW00}.
If integer $k$ is known from the context or irrelevant for the discussion, then we sometimes
use the simplified notion {\em expression-tree} for the notion $k$-expression-tree.

The following example shows that every clique $K_n$, $n \ge 1$,
has clique-width 2 and  NLC-width 1  and that every path $P_n$ has  clique-width at most 3 and NLC-width at most 3.

\begin{example} ~
\label{ex-cw}
\begin{enumerate}
\item[(1.)]
Every clique $K_n=(\{v_1,\ldots,v_n\},\{\{v_i,v_j\}~|~1\leq i<j\leq n\})$, $n \ge 2$,
has clique-width 2, by the following recursively defined expressions $X_{K_n}$.
\[
\begin{array}{lcl}
X_{K_2} & := &\eta_{1,2} (\bullet_1 \oplus \bullet_2)\\
X_{K_n} & := &\eta_{1,2} (\rho_{2\to 1}(X_{K_{n-1}})\oplus \bullet_2), {\rm if} ~n\ge 3
\end{array}
\]

\item[(2.)]
Every path $P_n=(\{v_1,\ldots,v_n\},\{\{v_1,v_2\},\ldots, \{v_{n-1},v_n\}\})$
has clique-width at most 3,  by the following recursively defined expressions $X_{P_n}$.
\[
\begin{array}{lcl}
X_{P_3} & := &\eta_{2,3}(\eta_{1,2}(\bullet_1 \oplus \bullet_2) \oplus \bullet_3)\\
X_{P_n} & := &\eta_{2,3} ( \rho_{3\to 2}(\rho_{2\to 1}(X_{P_{n-1}}))\oplus \bullet_3),{\rm if}  ~n\ge 4
\end{array}
\]

\item[(3.)] 
Every clique $K_n$, $n \ge 1$,
has  NLC-width 1,  by the following recursively defined expressions $X_{K_n}$.
\[
\begin{array}{lcl}
X_{K_1} & := & \bullet_1\\
X_{K_n} & := & X_{K_{n-1}}\times_{\{(1,1)\}} \bullet_1,{\rm if} ~n\ge 2
\end{array}
\]

\item[(4.)] 
Every path $P_n$
has NLC-width at most 3,  by the following recursively defined expressions $X_{P_n}$.
\[
\begin{array}{lcl}
X_{P_3} & := &(\bullet_1 \times_{\{(1,2)\}} \bullet_2) \times_{\{(2,3)\}}\bullet_3\\
X_{P_n} & := &\circ_{\{(1,1),(2,1),(3,2)\}}(X_{P_{n-1}})\times_{\{(2,3)\}}\bullet_3, {\rm if}~n\ge 4 
\end{array}
\]
\end{enumerate}
\end{example}

Next we give some examples for graph classes of bounded clique-width.
Distance hereditary graphs have clique-width at most 3 \cite{GR00}. 
Co-graphs, i.e. $P_4$-free graphs have clique-width at most 2 \cite{CO00}. Further,
many graph classes defined by a limited number of $P_4$ have bounded
clique-width, e.g. $P_4$-reducible graphs, $P_4$-sparse graphs, $P_4$-tidy, and $(q,t)$-graphs \cite{CMR00,MR99}.
A recent survey on graph classes of bounded clique-width is given in \cite{KLM07}.

On the other hand, the clique-width of permutation graphs, interval graphs, grids and planar graphs is
not bounded \cite{GR00}.

\subsection{Relations between graph parameters\label{sec-rel}}

Next we briefly survey the relation between tree-width, clique-width, and NLC-width.

\begin{theorem}[\cite{Joh98}]
\label{Tnlcwcw}
Every graph of clique-width $k$ has NLC-width at most $k$, and every graph
of NLC-width at most $k$ has clique-width at most  $2k$.
\end{theorem}

Thus we conclude that a set of graphs has bounded  clique-width if and only if 
it has bounded NLC-width. Both concepts are useful, because it is sometimes
much more comfortable to use NLC-width expressions instead of clique-width expressions and vice versa, respectively, see Chapter \ref{ch10}.

It is well known that every graph of bounded tree-width also has
bounded clique-width, see \cite{CO00,CR05,Wan94}. The best known bound 
is the following one shown by Corneil and Rotics.

\begin{theorem}[\cite{CR05}]
\label{Ttwcw}
Let $G$ be a graph of tree-width $k$, then $G$  has clique-width at most $3\cdot 2^{k-1}$.
\end{theorem}

Conversely, the tree-width of a graph can not be bounded in its  clique-width in general.
This shows e.g. the set of all complete graphs
($K_n$ has clique-width 2 and tree-width $n-1$). 
Under the additional assumption that we restrict to graphs that do not contain
arbitrary large complete bipartite graphs $K_{n,n}$, the tree-width of a graph can be bounded in its  clique-width \cite{GW00}.
Thus, if we restrict to graphs 
of bounded vertex degree or planar graphs, a set of graphs has bounded tree-width or bounded 
path-width if and only if 
it has bounded clique-width or bounded linear clique-width, respectively.

A further very useful and interesting relation between tree-width and clique-width has been shown using the concept of line graphs. A set of graphs has bounded tree-width  if and only if
the corresponding set of line graphs has bounded clique-width \cite{GW07b}.

\subsection{Definitions of basic graph parameters\label{ch-probl}}

Next we give definitions for the four basic graph 
parameters independence number, clique number,
chromatic number, and clique covering number.

\begin{problem}[Independent Set, {[GT20]} in \cite{GJ79}]~  \label{max-is}
\begin{description} \label{IS}
\item[Instance:] A graph $G=(V_G,E_G)$ and a positive integer $s \leq |V_G|$.
\item[Question:] Is there an independent set of size at least $s$ in $G$, i.e. a subset
$V'\subseteq V_G$, such that $|V'|\ge s$ and no two vertices of $V'$ are joined by an edge in $E_G$?
\end{description}
\end{problem}

The maximum value $s$ such that $G$ has an independent set of size $s$ is denoted as 
the {\em independence number} of graph $G$, denoted by $\alpha(G)$.

\begin{problem}[Clique, {[GT19]} in \cite{GJ79}]~ \label{max-cl}
\begin{description} \label{CL}
\item[Instance:] A graph $G=(V_G,E_G)$ and a positive integer $s \leq |V_G|$.
\item[Question:] Is there a clique of size at least $s$ in $G$, i.e. a subset
$V'\subseteq V_G$, such that $|V'|\ge s$ and every two vertices of $V'$ are joined by an edge in $E_G$?
\end{description}
\end{problem}

The maximum value $s$ such that $G$ has  a clique of size $s$ is denoted as the {\em clique number of $G$},
denoted by $\omega(G)$.

\begin{problem}[Partition Into Independent Sets, {[GT4]} in \cite{GJ79}]~  
\begin{description} \label{p-is}
\item[Instance:] A graph $G=(V_G,E_G)$ and a positive integer $s \leq |V_G|$.
\item[Question:] Is there a partition of $V_G$ into $s$ disjoint sets $V_1,\ldots,V_s$
such that $V_1 \cup \cdots \cup V_s=V_G$ and no set $V_t$, $1 \leq t \leq s$,
has two adjacent vertices?
\end{description}
\end{problem}

The minimum value $s$ such that $G$ has a partition into $s$ independent sets
is denoted as the {\em chromatic number of $G$}, denoted by $\chi(G)$.
Equivalently and motivating the notation chromatic number, $\chi(G)$ is the least integer $s$, such that
there is a {\em vertex coloring} $\col:V_G\to \{1,\ldots,s\}$ such that for every pair of adjacent vertices $v_1,v_2\in V_G$, $v_1\neq v_2$, it holds $\col(v_1)\neq \col(v_2)$.

\begin{problem}[Partition Into Cliques, {[GT15]} in \cite{GJ79}]~ 
\begin{description} \label{p-cl}
\item[Instance:] A graph $G=(V_G,E_G)$ and a positive integer $s \leq |V_G|$.
\item[Question:] Is there a partition of $V_G$ into $s$ disjoint sets $V_1,\ldots,V_s$
such that $V_1 \cup \cdots \cup V_s=V_G$ and every set $V_t$, $1 \leq t \leq s$, induces a complete subgraph?
\end{description}
\end{problem}

The minimum value $s$ such that $G$ has a partition into $s$ cliques
is denoted as the {\em clique covering number of $G$}, denoted by $\theta(G)$. 

\bigskip
The graph parameters  $\alpha$, $\omega$, $\chi$, and $\theta$  play an important rule
in the field of the research of perfect graphs. One of the most famous characterizations
for these graphs is that a graph $G$ is perfect if and only if for every induced subgraph $H$ of $G$ it holds $\omega(H) = \chi(H)$, if and only if for every induced subgraph $H$ of $G$ it holds $\alpha(H) = \theta(H)$.
Examples for perfect graph classes are bipartite graphs, chordal graphs, and co-graphs, see \cite{Hou06}
for an overview.
A further characterization for perfect graphs is that a graph $G$ is perfect if and only if $G$ contains no $C_{2n+1}$ and no $\overline{C_{2n+1}}$ as an induced subgraph.
Since the cycle on 5 vertices has tree-width 2 and clique-width 3, we conclude that
graphs of tree-width at most $k$ and graphs of clique-width at most $k$
are not perfect for every integer $k\ge 2$ and $k\ge 3$, respectively.

\bigskip
In Table \ref{table1} we survey the results of this paper.

\begin{table}
\begin{center}
\begin{tabular}{|l||lr|lr|lr|lr|}
\hline
\backslashbox{graph class}{parameter} &  \multicolumn{2}{c|}{$\alpha$} &  \multicolumn{2}{c|}{$\omega$} &  \multicolumn{2}{c|}{ $\chi$} & \multicolumn{2}{c|}{$\theta$} \\
\hline
trees/forest      & Lin   & Thm \ref{tree-in}& Lin  & folk  & Lin  & folk   &  Lin   & Thm \ref{tree-cov}        \\
$\TW_k$    & Lin   & Thm \ref{twmis} & Lin  & Thm \ref{twmc}  & P  & Thm  \ref{twcn} & P  &  Thm  \ref{twvcn}    \\
co-graphs         & Lin   & Thm \ref{co-in}  & Lin  & Thm \ref{co-cn}  & Lin  & Thm \ref{co-chrn}  & Lin  &  Thm \ref{co-clcovn} \\
$\CW_k$  & Lin   & Thm \ref{nlcmis}           & Lin  & Thm  \ref{nlcmc}  & P    & Thm \ref{nlcchr}  & P    &  Thm \ref{nlcccn}  \\
\hline
\end{tabular}
\end{center}
\caption{Overview on the time complexity of computing $\alpha$, $\omega$, $\chi$, and $\theta$ on special graph classes.} \label{table1}
\end{table}

\section{Tree-width and  polynomial time algorithms\label{ch9}}

\subsection{A general framework}

In order to solve hard problems restricted to graph classes of bounded tree-width, we will
perform a dynamic programming scheme on the
tree decomposition introduced in Definition \ref{D0}. 

Although computing the tree-width of a given graph is NP-complete \cite{ACP87}, for every fixed
integer $k$, the problem to decide whether a given graph $G$ has tree-width
at most $k$ can be solved in linear time and in the case of a positive answer
a tree decomposition of width $k$ for $G$ can be found in the same time \cite{Bod96}. 
For the purpose of convenience we want to restrict our algorithms to special binary decompositions
which is always possible by the following theorem.

\begin{theorem}[\cite{Klo94}] \label{klo} Let $G$ be a graph of tree-width $k$. Then $G$ has
a tree decomposition $({\mathcal X}=\{X_u \mid ~u\in 
V_T\},~T=(V_T,E_T))$  of width $k$, such that a root $r$ of $T$ can be chosen
such that the following five conditions are fulfilled.
\begin{enumerate}
\item Every node of $T$ has at most two children.
\item If a node $u$ of $T$ has two children $v$ and $w$, then $X_u=X_v=X_w$. In this case $u$
is called a {\em join node}.
\item If a node $u$ of $T$ has one child $v$, then one of the following tow conditions
hold true.
\begin{enumerate}
\item $|X_u|=|X_v|+1$ and $X_v\subset X_u$. In this case $u$
is called an {\em introduce node}.
\item $|X_u|=|X_v|-1$ and $X_u\subset X_v$. In this case $u$
is called a {\em forget node}.
\end{enumerate} 
\item If a node $u$ is a leaf of $T$, then $|X_u|=1$.
\item $|V_T|\in O(k\cdot |V_G|)$.
\end{enumerate}
\end{theorem}

A tree decomposition which fulfills the five conditions of Theorem \ref{klo}
is called a {\em nice tree decomposition} and can be found in linear time \cite{Klo94}. 
Let $G$ be a graph of tree-width $k$ and $({\mathcal X}=\{X_u \mid u\in 
V_T\},~T=(V_T,E_T))$ tree decomposition with root $r$ for $G$. 
For some node $u$ of $T$ we define $T_u$ as the subtree of $T$ rooted at $u$ and by ${\mathcal X}_u$ 
the set of all $X_v$, $v\in V_{T_u}$. Further by $G_u$ we define the subgraph of $G$ 
which is defined by all nodes in sets $X_v$ where $v=u$ or $v$ is a child of $u$ in $T$,
i.e. $G_u$ is defined by tree decomposition $({\mathcal X}_u,T_u)$.
The sets $X_u\in{\mathcal X}$ will be denoted  as {\em bags}.

\begin{figure}[ht]
\centerline{\epsfig{figure=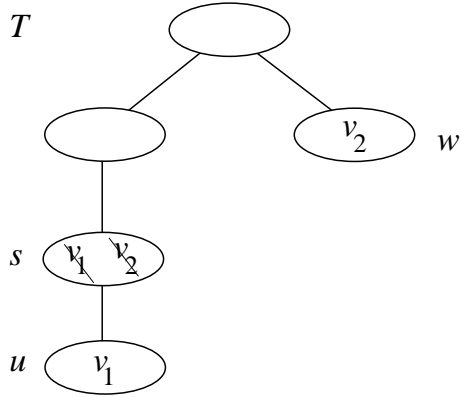,width=6cm}}
\caption[Illustration of a separator property of graphs defined by
tree decompositions]{\label{sep-pro}Separator property of some graph $G$ defined by
tree decompositions $({\mathcal X},T)$. Let $u, w \in V_T$ and $v_1\in X_u$,  $v_2\in X_w$ be two vertices of graph $G$. If $v_1\not\in X_s$ and $v_2\not\in X_s$, we know that vertices $v_1$ and $v_2$ are not adjacent in $G$.}
\end{figure}

Our solutions are based on a {\em separator property} of the vertices of graphs
given by a tree decomposition  $({\mathcal X},T)$ of width at most $k$. Let $u$, $w$ be two nodes of $T$ and $v_1\in X_u$,  $v_2\in X_w$ two vertices of the graph $G$ defined by decomposition $({\mathcal X},T)$. If there exists
some node $s$ of $T$ on the path from $u$ to $w$ in $T$ such that $v_1\not\in X_s$ and $v_2\not\in X_s$, then
$v_1$ and $v_2$ are not adjacent in $G$, see Fig. \ref{sep-pro}. Thus the at most $k+1$ vertices of bag $X_s$ separate the vertices in bags below $X_s$ from the remaining vertices of $G$.

In order to solve graph problems on tree-width bounded graphs we will use
the following bottom up dynamic programming scheme.

\begin{theorem}\label{schemeTW}
Let $\Pi$ be a graph problem and $k$ be a positive integer.
If there is a mapping $F$ that maps every  tree-de\-com\-po\-sition $({\mathcal X}=\{X_u \mid ~u\in 
V_T\},~T=(V_T,E_T))$ with root $r$ of width $k$ onto some structure $F(r)$, such that for all nodes
$v$,$w$ of $T$,
\begin{enumerate}
\item the size of $F(v)$ is polynomially bounded in the size of $({\mathcal X}_v,T_v)$,
\item the answer to $\Pi$ for $G_v$ is computable in polynomial time from $F(v)$,
\item for every leaf $u$ of $T$ structure $F(u)$ is computable in time $O(1)$,
\item for every join node $u$ with children $v,w$ structure $F(u)$ is computable in polynomial time from $F(v)$ and $F(w)$, and
\item for every introduce node and every forget node $u$ with child $v$ structure $F(u)$ is computable in polynomial time from $F(v)$.
\end{enumerate}
Then for every decomposition $({\mathcal X},T)$ of width $k$, the answer to $\Pi$ for graph $G_r$ is computable in polynomial time from decomposition $({\mathcal X},T)$. 
\end{theorem}

There are further dynamic programming approaches to solve hard problems on tree-width bounded graphs. For
example in \cite{AP89} the perfect elimination order of the vertices of a partial $k$-tree is used to
solve hard problems on tree-width bounded graphs.

\subsection{Computing $\alpha$, $\omega$, $\chi$, and $\theta$ on graphs of bounded tree-width\label{stw}}

We next apply the general scheme of Theorem \ref{schemeTW} for computing the four basic graph
parameters $\alpha$, $\omega$, $\chi$, and $\theta$ on graphs of bounded tree-width
in polynomial time.

\subsubsection{Independence number}

First we consider the problem of finding the size of a maximum independent set (Problem \ref{max-is}) in a graph given by some tree decomposition.

Let  $({\mathcal X}=\{X_u \mid u\in 
V_T\},~T=(V_T,E_T))$ be a tree decomposition for some graph $G$ of width $k$ with root $r$.
For every node $u$ of $T$ we define a $2^{k+1}$-tuple $F(u)$ which contains for every subset $X\subseteq X_u$ 
an integer $a_X$, i.e. $F(u)=(a_X~|~X\subseteq X_u)$. The value of $a_X$ denotes the size
of a maximum independent set $U\subseteq V_{G_u}$ in graph $G_u$ such that $U \cap X_u = X$.
Note that because of our separator property, vertices from $U-X$ will not get any further edges in bag $X_u$ or
some bag $X_w$ for some node $w$ which is not a child of $u$ in $T$.

Then $F(r)$ is bounded in $k$ independently of the size of $({\mathcal X},T)$, because by the definition every 
bag contains at most $k+1$ vertices and thus $F(r)$ has at most $2^{k+1}$ entries. The following observations 
show that for every leaf $u$ of $T$ structure $F(u)$ is
computable in time $O(2^{k+1})$, for every  join node $u$ with children $v$,$w$ structure $F(u)$ is computable in time $O(2^{k+1})$ from $F(v)$ and $F(w)$,
and for every introduce node and every forget node $u$ with child $v$ structure $F(u)$ is computable in time $O(2^{k+1})$ from $F(v)$.

\begin{enumerate}
\item If $u$ is a leaf of $T$, such that $X_u=\{v_1\}$ for some $v_1\in V_G$.
We define $F(u):=(a_{\emptyset}=0, a_{\{v_1\}}=1)$.

\item If $u$ is a join node with children $v,w$ of $T$. 
Let $F(v)=(a_X~|~X\subseteq X_v)$, $F(w)=(b_X~|~X\subseteq X_w)$, and $i_X$ be the size of the largest independent set in $X\subseteq X_u$. 
Then we define $F(u):=(c_X~|~X\subseteq X_u)$, where $\forall X \subseteq X_u$, $c_X:=a_X+b_X-i_X$.

\item If $u$ is an introduce node with child $v$ of $T$, such that $X_u-X_v=\{v'\}$ for some $v'\in V_G$. 
Let $F(v)=(a_X~|~X\subseteq X_v)$. 
Then we define $F(u)=(b_X~|~X\subseteq X_u)$, where $\forall X \subseteq X_v$, 
$b_X:=a_X$ and
\[b_{X\cup \{v'\}} := \left\{\begin{array}{ll}
a_X+1  & \text{if $v'$ is not adjacent to some vertex from } X\\
-\infty  & \text{else } \\
\end{array}\right.\]

\item If $u$ is a forget node with child $v$ of $T$, such that $X_v-X_u=\{v'\}$ for some $v'\in V_G$.
Let $F(v)=(a_X~|~X\subseteq X_v)$.
Then we define $F(u):=(b_X~|~X\subseteq X_u)$, where $\forall X \subseteq X_u$, $b_X:=\max\{a_X,a_{X\cup\{v'\}}\}$.
\end{enumerate}

After a dynamic programming computation of $F(r)$ we can easily compute the size of a maximum independent set in graph $G$ by $\alpha(G):=\max_{a\in F(r)}a$

\begin{theorem} \label{twmis}
The independence number of a graph of
bounded tree-width can be computed in linear time.
\end{theorem}

In \cite{Chl02} it is shown that for every graph $G$ the value of $|V_G|-\tw(G)$
always is an upper bound for the independence number $\alpha(G)$.

\subsubsection{Clique number}

Next we consider the problem of finding the size of a maximum clique (Problem \ref{max-cl}) in
a graph given by some tree decomposition.

Let $G$ be some graph and $({\mathcal X}=\{X_u \mid ~u\in 
V_T\},~T=(V_T,E_T))$ be a tree decomposition of width $k$ for $G$. In order
to compute the value of $\omega(G)$ obviously a similar solution as given for independent set problem 
above is possible. Alternatively one could use the well known result that for every clique $C=(V_C,E_C)$ in graph $G$
there exists some bag $X_u\in{\mathcal X}$ such that $V_C\subseteq X_u$, see \cite{BM93}. This allows us to compute the value of $\omega(G)$ for graph $G$ of tree-width $k$ by 
\[\omega(G):=\max_{u\in V_T} \max_{C\subseteq X_u \atop {G[C] \text{~clique}}} |C|.\]

\begin{theorem} \label{twmc}
The clique number of a graph of bounded tree-width can be computed in linear time.
\end{theorem}

\subsubsection{Chromatic number\label{sec-chrn}}

Further we consider the problem of finding the minimum number of independent sets (Problem \ref{p-is})
covering a graph given by some tree decomposition.

Let  $({\mathcal X}=\{X_u \mid ~u\in 
V_T\},~T=(V_T,E_T))$ be a tree decomposition for some graph $G$ of width $k$ with root $r$.
For every node $u$ of $T$, we define a set $F(u)$ which contains for every partition of $V_{G_u}$ into independent
sets  $V_1,\ldots,V_s$  a $2^{k+1}$-tuple $t=(\ldots,a_{X},\ldots,a)$ which contains for every nonempty subset $X\subseteq X_u$ a boolean value $a_X$ and one integer  $a$.  For some disjoint partition $V_1,\ldots,V_s$  of $V_{G_u}$ into independent sets, the value of $a_X$ is $1$, if and only if $V_i\cap X_u=X$ for some $1\leq i \leq r$  and the value of $a$ denotes the number of independent sets $V_i$ such that $V_i\cap X_u=\emptyset$.

Then $F(r)$ is polynomially bounded in the size of $({\mathcal X},T)$, because 
every element of $F(r)$ has $2^{k+1}$ entries, $2^{k+1}-1$ from $\{0,1\}$ and one from $\{1,\ldots,|V_G|\}$, i.e.
$|F(r)|\leq 2^{2^{k+1}-1}\cdot |V_G|$. The following observations 
show that for every leaf $u$ of $T$ structure $F(u)$ is
computable in time $O(1)$, for every  join node $u$ with children $v$,$w$ structure $F(u)$ is computable in polynomial time from $F(v)$ and $F(w)$,
and for every introduce node and every forget node $u$ with child $v$ structure $F(u)$ is computable in polynomial time from $F(v)$.

\begin{enumerate}
\item  If $u$ is a leaf of $T$, such that $X_u=\{v_1\}$ for some $v_1\in V_G$.
We define  $F(u):=\{(a_{\{v_1\}}=1,a=0)\}$.

\item If $u$ is a join node with children $v,w$ of $T$. 
Then we define $F(u):=\{ (\ldots,a_{X},\ldots,a+b) ~|~ (\ldots,a_{X},\ldots,a) \in F(v), (\ldots,b_{X},\ldots,b) \in F(w), a_{X}=b_{X}, \text{for all } X\subseteq X_u \}$.

\item  If $u$ is an introduce node with child $v$ of $T$, such that $X_u-X_v=\{v'\}$ for some $v'\in V_G$.
We consider every partition of $X_v$ into independent sets in order to insert a new independent set which just
contains vertex $v'$ and to extend one existing independent set of a partition of $X_v$ by vertex $v'$.

Thus we define $F(u)$ as follows. For every tuple $t\in F(v)$ we insert a tuple $t'$ into $F(u)$ which contains
the same values as $t$ and additionally $a_{\{v'\}}:=1$. Further for every tuple $t\in F(v)$ and every $X\subseteq X_v$
such that $a_X=1$ in $F(v)$ and $X\cup\{v'\}$ is an independent set of graph $G$, we insert a tuple $t'$ into $F(u)$ which contains the same values as $t$ but  $a_X:=0$ and additionally $a_{X\cup\{v'\}}:=1$.
In both cases the value of $a$ of $t'$ is the same as in $t$.


\item  If $u$ is a forget  node with child $v$ of $T$, such that $X_v-X_u=\{v'\}$ for some $v'\in V_G$.
We know that in every partition of  $X_v$ there is exactly one independent set $X$ which contains $v'$.
If $X$ does not contain any further vertex, we have to increase the value of $a$ in $F(u)$ by one, otherwise
we know that $X-\{v'\}$ is an independent set in $G_u$, i.e. $a_{X-\{v'\}}=1$ in $F(u)$.

Thus we define $F(u)$ as follows. For every tuple $t=(\ldots,a_{X},\ldots,a) \in F(v)$ we insert a tuple $t':=(\ldots,b_{X},\ldots,b)$ into $F(u)$ which is defined as follows. If  $a_{\{v'\}}=1$, then we 
define for every $X\subseteq X_v$ $b_{X-\{v'\}}:=a_X$ and $b:=a+1$. Otherwise (i.e. $a_{\{v'\}}=0$), we
define for every $X\subseteq X_v$ $b_{X-\{v'\}}:=a_X$ and $b:=a$.
\end{enumerate}
Note that in all four cases set $F(u)$ has at most $2^{|X_u|}$ entries.

After a dynamic programming computation of $F(r)$ we can  compute the chromatic number of 
graph $G$ by $\chi(G):=\min_{t \in F(r)} \sum_{a\in t} a$.

\begin{theorem} \label{twcn}
The chromatic number of a graph of
bounded tree-width can be computed in polynomial time.
\end{theorem}

In \cite{Chl02} it is shown that for every graph $G$ the value of $\tw(G)+1$
is always an upper bound for the chromatic number $\chi(G)$.

If we consider Problem \ref{p-is} for the case that we look for a partition 
into a minimum number of independent edge sets, we obtain the graph parameter chromatic index, see \cite{Viz64}, \cite{Gup66}, which has
been shown to be computable in linear time on graphs of
bounded tree-width in \cite{ZFN00}.

\subsubsection{Clique covering number}

Further we consider the problem of finding the minimum number of cliques  (Problem \ref{p-cl}) covering a graph given by some tree decomposition.

Let  $({\mathcal X}=\{X_u \mid ~u\in V_T\},~T=(V_T,E_T))$ be a tree decomposition for some graph $G$ of width $k$ with root $r$. We will proceed similarly to the solution
given in Section \ref{sec-chrn} for computing the chromatic number.
For every node $u$ of $T$, we define a set $F(u)$ which contains for every disjoint partition of $V_{G_u}$ into cliques $V_1,\ldots,V_s$ a $2^{k+1}$-tuple $t=(\ldots,a_{X},\ldots,a)$ which contains for every nonempty subset $X\subseteq X_u$ a boolean value $a_X$ and one integer  $a$.  For some disjoint partition $V_1,\ldots,V_s$ of $V_{G_u}$ into cliques, the value of $a_X$ is $1$, if and only if  $V_i=X$ for some $1\leq i \leq r$ and the value of $a$ denotes the number of cliques $V_i$ such that $V_i\cap X_u=\emptyset$.

Then $F(r)$ is polynomially bounded in the size of $({\mathcal X},T)$, because $|F(r)|\leq 2^{2^{k+1}-1}\cdot |V_G|$. Further for every leaf $u$ of $T$ structure $F(u)$ is
computable in time $O(1)$, for every  join node $u$ with children $v$,$w$ structure $F(u)$ is computable in polynomial time from $F(v)$ and $F(w)$,
and for every introduce node and every forget node $u$ with child $v$ structure $F(u)$ is computable in polynomial time from $F(v)$. This follows by step (1) and (2) given
in Section \ref{sec-chrn}  and by replacing {\em independent set} by {\em clique} in step (3) and (4) given in Section \ref{sec-chrn}.

After a dynamic programming computation of $F(r)$ we can compute the  clique covering number
of graph $G$ by $\theta(G):=\min_{t \in F(r)} \sum_{a\in t} a$

\begin{theorem} \label{twvcn}
The clique covering number of a graph of
bounded tree-width can be computed in polynomial time.
\end{theorem}

\subsection{Computing  $\alpha$, $\omega$, $\chi$, and $\theta$ on trees}

We next show that for graphs of tree-width one, i.e. for trees, our shown algorithms 
can be simplified.

\subsubsection{Independence number \label{ssist}}

Let $T$ be some tree. The independence number $\alpha(T)$ can be computed by
$\alpha(T)=\alpha(T_r)$, where $T_r$ is the corresponding rooted tree
by choosing an arbitrary vertex $r$ of $T$ as a root. The value of $\alpha(T_r)$ (and thus $\alpha(T)$) can
be  computed by dynamic programming as follows.

\begin{enumerate}
\item If $|V_{T_r}|=1$, i.e. $r$ is a leaf of $T_r$, then $\alpha(T_r):=1$.

\item If $|V_{T_r}|\ge 2$,  i.e. $r$ is an inner node  of $T_r$, then

\[\alpha(T_r):=\max\{ \sum_{\mbox{$c$ child of $r$}} \alpha(T_{c}), ~1+ \sum_{\mbox{$g$ grandchild of $r$}} \alpha(T_{g})\}\]
\end{enumerate}

\begin{theorem} \label{tree-in}
For every tree its independence number can be computed in linear time.
\end{theorem}

\subsubsection{Clique number}

Obviously for every tree $T$,
if $|V_T|=1$, then $\omega(T):=1$ and if $|V_T|\ge 2$, then $\omega(T):=2$.

\subsubsection{Chromatic number}

Obviously for every tree $T$,
if $|V_T|=1$, then $\chi(T):=1$ and if $|V_T|\ge 2$, then $\chi(T):=2$, since
every tree is a bipartite graph.

\subsubsection{Clique covering number}

Since trees are perfect, we know that  for every induced subgraph $H$ of some
tree $T$ it holds $\theta(H)=\alpha(H)$, and thus we can compute clique covering number $\theta(T)$ by the same algorithm as
shown for $\alpha(T)$ above.

\begin{theorem} \label{tree-cov}
For every tree its clique covering number can be computed in linear time.
\end{theorem}

\subsection{Complement problems\label{tw-cp}}

Let $\Pi$ be a decision problem for graphs. We define the corresponding {\em complement problem} $\overline{\Pi}$ by
\[\overline{\Pi}:=\{\overline{G}~|~ G {\rm ~satiesfies~} \Pi\}.\]

For several graph problems the corresponding complement problem is also of interest. For
example the complement problem of the independent set problem (Problem \ref{max-is}) is the clique problem (Problem \ref{max-cl})
and the complement problem of the partition into independent sets problem (Problem \ref{p-is})  is the partition into cliques problem (Problem \ref{p-cl}).

Since for some set of graphs ${\mathcal L} \subseteq \TW_k$, the corresponding set of complement graphs
$\overline{\mathcal L}:=\{\overline{G}~|~G\in{\mathcal L} \}$ not necessarily has bounded tree-width, the solvability
of complement problems on tree-width bounded graphs  are worthwhile to research.
In \cite{GKS00} Gupta et al. give a logical framework for solving complement problems on tree-width bounded graphs in polynomial time.

\subsection{Tree-width and monadic second order logic}

On graph classes of bounded tree-width, all graph properties and optimization problems which are expressible in monadic second order logic with quantifications over vertices, vertex sets, edges, and edge sets ($\MSOB$-logic) are decidable in linear time \cite{CM93}. This implies the existence of linear time algorithms for computing the independence number and  clique number on graphs of bounded tree-width. Note that the problems partition into independent sets and partition into cliques are not expressible in $\MSOB$-logic.

\section{Clique-width and  polynomial time algorithms\label{ch10}}

\subsection{A general framework\label{sec_fra}}

In order to solve hard problems restricted to graph classes of bounded clique-width,
we recall a dynamic programming approach on the tree structure of a clique-width or an NLC-width expression from \cite{EGW01a}.

Recently it has been shown that computing the clique-width and NLC-width of a given graph is NP-hard \cite{FRRS06,GW05}.
For every fixed integer $k\le 3$ or $k\le 2$, the problem to decide whether a given graph has clique-width
at most $k$ or NLC-width at most $k$ can be solved in polynomial time and in the case of a positive answer
a $k$-expression can be constructed in the same time \cite{CPS85,CHLRR00,LMR07}.
For every fixed integer $k\ge 4$ or $k\ge 3$, the problem to decide whether a given graph has clique-width
at most $k$ or NLC-width at most $k$ is still open.
Nevertheless we can use the approximations for rank-width shown by Oum and Seymour in \cite{OS06,Oum05,Oum06} in order to obtain 
approximations for clique-width and a corresponding clique-width expression. The best known 
result is the following.

\begin{theorem}[\cite{Oum06}] \label{To2}
For every fixed integer $k$ there is a $O(|V_G|^3)$ algorithm that either outputs a clique-width $(8^{k}-1)$-expression of an input graph $G$, or confirms that the clique-width of $G$ is larger that $k$.
\end{theorem}

Every clique-width $k$-expression can be transformed into an equivalent NLC-width $k$-expression within linear time \cite{Joh98}. Thus, the last theorem
implies that for every fixed integer $k$, for every set ${\mathcal L} \subseteq \CW_k$ and every set ${\mathcal L} \subseteq \NLC_k$,  we can assume within cubic time every graph $G\in{\mathcal L}$
to be given with some $(8^{k}-1)$-expression.

For some node $u$ of expression-tree $T$, let $T(u)$ be the subtree of $T$ rooted at $u$.
Note that tree $T(u)$ is always an expression-tree. The expression $X(u)$ defined by
$T(u)$ can simply be determined by traversing the tree $T(u)$ starting from the root, where the left children are visited first. $X(u)$ defines a (possibly) relabeled induced subgraph $G[u]$ of $G$.

\begin{figure}[ht]
\centerline{\epsfig{figure=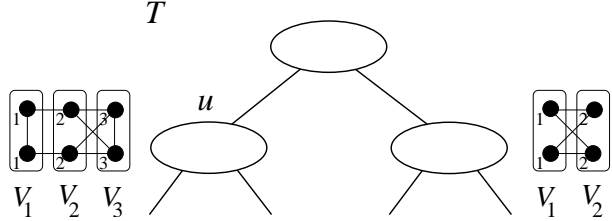,width=8cm}}
\caption[Illustration of a separator property of graphs defined by 
$k$-expressions]{\label{neigh-pro}Neighbourhood property of graphs defined by
$k$-expression tree $T$. For every node $u\in V_T$, the vertices of subgraph $G[u]$ can be divided with respect to their labels into at most $k$ modules $V_1,\ldots, V_k$ in the graph 
defined by $T$.}
\end{figure}

Our solutions are based on a {\em neighbourhood property} of the vertices of graphs
given by a $k$-expression $X$ defining a corresponding $k$-expression tree $T$. For every node $u$ of $T$,  the vertices
of subgraph $G[u]$ form a $k$-module, i.e.  every set $V_i=\{\lab(v)=i~|~ v\in V_{G[u]}\}$, $1\leq i\leq k$, is a
module of $G[V_G-(V_1\cup\ldots\cup V_k)\cup V_i]$. That is, all vertices in set $V_i$, $1\leq i \leq k$, will be treated equally by all
operations in $T$ on the path from $u$ to the root of $T$, see Fig. \ref{neigh-pro}.

The tree structure of such  $k$-expressions can be used to solve hard problems by the
following  general bottom up dynamic programming scheme.

\begin{theorem} [\cite{EGW01a}] \label{Tsche1}
Let $\Pi$ be a graph problem and $k$ be a positive integer.
If there is a mapping $F$ that maps each clique-width $k$-expression $X$
onto some structure $F(X)$,
such that for all clique-width $k$-expressions $X,Y$ and all $a,b\in[k]$
\begin{enumerate}
\item the size of $F(X)$ is polynomially bounded in the size of $X$,
\item the answer to $\Pi$ for $\graph(X)$ is computable in polynomial time from $F(X)$,
\item $F(\bullet_a)$  is computable in time $O(1)$,
\item $F(X \oplus Y)$ is computable in polynomial time from $F(X)$ and $F(Y)$, and
\item $F(\eta_{a,b}(X))$ and $F(\rho_{a \to b}(X))$ are computable in polynomial time from $F(X)$.
\end{enumerate}
Then for every clique-width $k$-expression $X$, the answer to $\Pi$ for graph $\graph(X)$ is computable in polynomial time from expression $X$. 
\end{theorem}

Theorem \ref{Tsche1} also works for NLC-width $k$-expressions built with
the operations $\bullet_a$, $\times_S$, and $\circ_R$ instead of
$\bullet_a$, $\oplus$, $\eta_{a,b}$, and $\rho_{a \to b}$. In this case $F(X \times_S Y)$ has to be
computable in polynomial time from $F(X)$ and $F(Y)$, and $\circ_R(X)$ has to be computable
in polynomial time from $F(X)$.

The given dynamic programming approach has been used in \cite{EGW01a},\cite{GW06}, \cite{Gur07e}, \cite{GK03},\cite{KR03},\cite{Rao07a} to solve a large number of NP-complete graph problems on graph classes of bounded clique-width.

\subsection{Computing $\alpha$, $\omega$, $\chi$, and $\theta$ on graphs of bounded clique-width}

We next give polynomial time algorithms using for computing the four basic graph
parameters  $\alpha$, $\omega$, $\chi$, and $\theta$  on graphs of bounded clique-width using the general scheme of Theorem \ref{Tsche1}.
For the sake of convenience and to emphasize the advantages of clique-width and NLC-width
operations, we will use clique-width expressions
for the problems  independent set and partition into independent sets and NLC-width expressions for the problems clique and partition into cliques.

\subsubsection{Independence number \label{cwis}}

Let us first consider the problem of computing the independence number 
(Problem \ref{max-is}) on graphs of bounded clique-width.

Let $G$ be a graph defined by some clique-width $k$-expression $X$.
Let $F(X)$ be the $(2^k-1)$-tuple
$(\ldots, a_{L}, \ldots)$, which contains for every $L\subseteq [k]$, $L\neq \emptyset$, a
non negative integer $a_{L}$, which denotes the size of a largest
independent set $U$ in graph $\val(X)$ such that $\{\lab(u)~|~u\in U\}=L$.

Then $F(X)$ is bounded in $k$ independently of the size of $X$, because 
$F(X)$ has at exactly $2^k-1$ entries. The following observations 
show that $F(\bullet_i)$ is
computable in time $O(2^k)$, $F(X \oplus Y)$ is computable in time  $O(2^k)$ from $F(X)$
and $F(Y)$, and $F(\eta_{i,j}(X))$ and $F(\rho_{i \to j}(X))$ are computable in time  $O(2^k)$ from $F(X)$.

\begin{enumerate}
\item We define $F(\bullet_i):=(\ldots, a_{L}, \ldots)$, where $\forall L \subseteq [k]$
\[a_L := \left\{\renewcommand{\arraystretch}{1.5}\begin{array}{ll}
1  & \text{if } L=\{i\}\\
0  & \text{if } L\neq\{i\}.\\
\end{array}\right.~\]

\item Let $F(X)=(\ldots, a_{L}, \ldots)$ and $F(Y)=(\ldots, b_{L}, \ldots)$, then we define
$F(X \oplus Y):=(\ldots, c_{L}, \ldots)$, where 
$c_L:=\max_{L=L_1 \cup L_2}a_{L_1}+b_{L_2}$, $L_1,L_2\subseteq[k]$.

\item Let $F(X)=(\ldots, a_{L}, \ldots)$, then we define $F(\eta_{i,j}(X)) := (\ldots, b_{L}, \ldots)$, where  $\forall L \subseteq [k]$
\[b_L := \left\{\renewcommand{\arraystretch}{1.5}\begin{array}{ll}
a_L  & \text{if } \{i,j\}\nsubseteq L\\
0  & \text{if } \{i,j\}\subseteq L\\
\end{array}\right.
\]

\item Let $F(X)=(\ldots, a_{L}, \ldots)$, then we define $F(\rho_{i \to j}(X)):=(\ldots, b_{L}, \ldots) $, where $\forall L \subseteq [k]$
\[b_L := \left\{\renewcommand{\arraystretch}{1.5}\begin{array}{ll}
a_L & \text{if } i \not\in L \text{ and } j \not\in L  \\
\max\{a_L,a_{L-\{j\}\cup\{i\}}\}  & \text{if } i \not\in L \text{ and } j \in L  \\
0  & \text{if } i\in L\\
\end{array}\right.
\]
Obviously in graph $\val(\rho_{i \to j}(X))$ there exists no vertex labeled by $i$, thus for every set $L$ with $i\in L$ we know that $b_L=0$.
\end{enumerate}

After a dynamic programming computation of $F(X)$ we can  compute the size of a maximum independent set in graph $\val(X)$
by $\alpha(\val(X)):=\max_{a\in F(X)}a$.

\begin{theorem}\label{nlcmis}
The independence number of a graph of bounded clique-width can be computed in linear time, if the
graph is given by some clique-width $k$-expression.
\end{theorem}

\subsubsection{Clique number\label{cwmaxcl}}

Let us next consider the problem of computing the  clique number (Problem \ref{max-cl}).

In order have some knowledge on the order of the operations\footnote{In order to handle clique-width expressions 
we want to mention the normal form for clique-width expressions defined in \cite{EGW03}.} in a given expression $X$, we assume that
$G$ is given by some some NLC-width $k$-expression $X$.
Let $F(X)$ be the $(2^k-1)$-tuple
$(\ldots, a_{L}, \ldots)$, which contains for every $L\subseteq [k]$, $L\neq \emptyset$, a
non negative integer $a_{L}$, which denotes the size of a largest
clique $C$ in graph $\val(X)$ such that $\{\lab(v)~|~v\in C\}=L$.

Then $F(X)$ is  bounded in $k$ independently from the size of $X$, because 
$F(X)$ has exactly $2^k-1$ entries. Next we will
show that that $F(\bullet_i)$ is
computable in time $O(1)$, $F(X \times_S Y)$ is computable in $O(2^k)$ time from $F(X)$
and $F(Y)$, and  $F(\circ_R(X))$ is computable in time $O(2^k)$ from $F(X)$.

\begin{enumerate}
\item We define $F(\bullet_i):=(\ldots, a_{L}, \ldots)$, where  $\forall L \subseteq [k]$
\[a_L := \left\{\renewcommand{\arraystretch}{1.5}\begin{array}{ll}
1  & \text{if } L=\{i\}\\
0  & \text{if } L\neq\{i\}.\\
\end{array}\right.\]

\item Let $F(X)=(\ldots, a_{L}, \ldots)$ and $F(Y)=(\ldots, b_{L}, \ldots)$,
then we define
$F(X \times_S Y):=(\ldots, c_{L}, \ldots)$, where the values for $c_L$ are defined as follows. 

We say relation $S=\{(i_1,j_1),\ldots,(i_l,j_l)\}$ {\em defines a join for
the label sets $L_1,L_2$, denoted by $L=L_1 \uplus L_2$}, if there is a subset $I\subseteq [l]$ such that 
$L_1=\cup_{i'\in I} i_{i'}$ and $L_2=\cup_{i'\in I} j_{i'}$ and $L=L_1 \cup L_2$.  Then we
can define $\forall L \subseteq [k]$
\[c_L := \left\{\renewcommand{\arraystretch}{1.5}\begin{array}{ll}
\max_{L=L_1\uplus L_2}\{a_{L_1}+b_{L_2},a_{L},b_{L}\} & \text{if there exists some } L_1 \uplus L_2=L   \\
\max\{a_{L},b_{L}\}  & \text{else  } \\
\end{array}\right.
\]

\item Let $F(X)=(\ldots, a_{L}, \ldots)$, then we define $F(\circ_R(X)) := (\ldots, b_{L}, \ldots)$, where $\forall L \subseteq [k]$
\[b_L := \left\{\renewcommand{\arraystretch}{1.5}\begin{array}{ll}
\max_{R(L')=L}a_{L'} & \text{if there exists some } L'\subseteq [k]: R(L')=L  \\
0  & \text{else } \\
\end{array}\right.
\]
\end{enumerate}

After a dynamic programming computation of $F(X)$ we  can compute the size of a maximum clique in graph $\val(X)$ by
$\omega(\val(X)):=\max_{a\in F(X)}a$.

\begin{theorem}\label{nlcmc}
The clique number of a graph of
bounded clique-width can be computed in linear time, if the
graph is given by some  $k$-expression.
\end{theorem}

\subsubsection{Chromatic number\label{ch534}}

Next we consider the problem of computing the chromatic number (Problem \ref{p-is}) on graphs of bounded clique-width.

Let $G$ be a graph given by  some clique-width $k$-expression $X$.
For a disjoint partition of $V_G$ into 
independent sets $V_1,\ldots,V_r$ let ${\mathcal M}$ be the multi set\footnote{A {\em multi set} is a set
that may have several equal elements. For a multi set with elements $x_1,\ldots,x_n$ we write
${\mathcal M}=\langle x_1,\ldots,x_n \rangle$.
There is no order on the elements of ${\mathcal M}$.
The number how often an element $x$ occurs in ${\mathcal M}$ is denoted by $\psi({\mathcal M},x)$.
Two multi sets ${\mathcal M}_1$ and ${\mathcal M}_2$
are {\em equal} if for each element $x \in {\mathcal M}_1 \cup {\mathcal M}_2$,
$\psi({\mathcal M}_1,x)=\psi({\mathcal M}_2,x)$, otherwise they are called
{\em different}. The empty multi set is denoted by $\langle \rangle$. The size of a multi set ${\mathcal M}$ is the
number of its elements, denoted by $|{\mathcal M}|$.}
$\langle \lab(V_1),\ldots,\lab(V_r) \rangle$.
Let $F(X)$ be the set of all mutually different multi sets
${\mathcal M}$ for all disjoint partitions of vertex set $V_G$ into independent sets.

Then $F(X)$ is polynomially bounded in the size of $X$, because 
$F(X)$ has at most $(|V_G|+1)^{2^k-1}$ mutually different multi sets each with
at most $|V_G|$ nonempty subsets of $[k]$. The following observations show that $F(\bullet_i)$ is
computable in time $O(1)$, $F(X \oplus Y)$ is computable in polynomial time from $F(X)$
and $F(Y)$, and $F(\eta_{i,j}(X))$ and $F(\rho_{i \to j}(X))$ are computable in polynomial time from $F(X)$.

\begin{enumerate}
\item We define $F(\bullet_i):=\{ \langle \{i\} \rangle \}$

\item
Starting with set $D:=\{ \langle \rangle \} \times F(X) \times F(Y)$
extend $D$ by all triples that can be obtained from some 
triple $({\mathcal M},{\mathcal M}',{\mathcal M}'') \in D$ by removing a set $L'$ from ${\mathcal M}'$
or a set $L''$ from ${\mathcal M}''$ and inserting it into ${\mathcal M}$, or
by removing both sets and inserting $L' \cup L''$ into ${\mathcal M}$.

We define $F(X \oplus Y):=\{ {\mathcal M} \mid ~ ({\mathcal M},\langle \rangle,\langle \rangle) \in D\}$.

$D$ gets at most $(|V_G|+1)^{3(2^k-1)}$ triples and thus is computable in polynomial time.

\item
We define $F(\eta_{i,j}(X)) := \{ \langle L_1, \ldots, L_r \rangle \in F(X) 
\mid  \{i,j\} \not\subseteq L_t ~ \text{for} ~ t=1,\ldots,r \}$.

\item
We define  $F(\rho_{i \to j}(X)) := \{ \langle \rho_{i \to j}(L_1), \ldots, \rho_{i \to j}(L_r) \rangle \mid  \langle L_1,\ldots,L_r \rangle \in F(X) \}$.

For a relabeling $\rho_{i \to j}$ let $R_{i \to j}:[k] \to [k]$
be defined by $R_{i \to j}(t):=t$ if $t \not= i$, and
$R_{i \to j}(t):=j$ if $t=i$. For $L \subseteq [k]$
let  $\rho_{i \to j}(L):=\{R_{i \to j}(t)~|~t\in L\}$.
\end{enumerate}

There is a partition of the vertex set of $\graph(X)$ into $r$ independent sets if and only if there is some ${\mathcal M} \in F(X)$ consisting of $r$ label sets. 
The chromatic number of graph  $\graph(X)$ can be obtained by
$\chi(\val(X)):=\min_{{\mathcal M} \in F(X)} |{\mathcal M}|$.

\begin{theorem} \label{nlcchr}
The chromatic number of a graph of
bounded clique-width can be computed in polynomial time.
\end{theorem}

The time complexity of computing the graph parameter chromatic index (minimum
number of colors needed to color the edges of a given graph) is open up to now
even for co-graphs.

\subsubsection{Clique covering number\label{cwccn}}

Finally we consider the problem of computing the clique covering number (Problem \ref{p-cl}) on graphs of bounded clique-width..

Let $G$ be a graph given by  some NLC-width $k$-expression $X$. 
For a disjoint partition of $V_G$ into cliques $V_1,\ldots,V_r$ let ${\mathcal M}$ be the multi set
$\langle \lab(V_1),\ldots,\lab(V_r) \rangle$.
Let $F(X)$ be the set of all mutually different multi sets
${\mathcal M}$ for all disjoint partitions of vertex set $V_G$ into cliques.

Then $F(X)$ is polynomially bounded in the size of $X$, because 
$F(X)$ has at most $(|V_G|+1)^{2^k-1}$ mutually different multi sets each with
at most $|V_G|$ nonempty subsets of $[k]$. The following observations 
show that $F(\bullet_i)$ is
computable in time $O(1)$, $F(X \times_S Y)$ is computable in polynomial time from $F(X)$
and $F(Y)$, and  $F(\circ_R(X))$ is computable in polynomial time from $F(X)$.

\begin{enumerate}
\item
We define $F(\bullet_i):=\{ \langle \{i\} \rangle \}$

\item In order to compute $F(X \times_S Y)$ from $F(X)$ and $F(Y)$ we start with set 
$D:=\{\langle \rangle \} \times F(X) \times F(Y)$ and
extend $D$ by all triples that can be obtained from some 
triple $({\mathcal M},{\mathcal M}',{\mathcal M}'') \in D$ by removing a set $L'$ from ${\mathcal M}'$
or a set $L''$ from ${\mathcal M}''$ and inserting it into ${\mathcal M}$, or if $S$ defines a join for
the label sets $L',L''$ (defined in Section \ref{cwmaxcl})
by removing both sets and inserting $L' \cup L''$ into ${\mathcal M}$.

We define $F(X \times_S Y):=\{ {\mathcal M} \mid ~ ({\mathcal M},\langle \rangle,\langle \rangle) \in D\}$.

$D$ gets at most $(|V_G|+1)^{3(2^k-1)}$ triples and thus is computable in polynomial time.

\item We define $F(\circ_R) := \{ \langle \circ_R(L_1), \ldots, \circ_R(L_r) \rangle 
\mid ~ \langle L_1,\ldots,L_r \rangle \in F(X) \}$.

For a relabeling $\circ_R$ and $L \subseteq [k]$
let  $\circ_R(L):=\{R(t)~|~t\in L\}$.
\end{enumerate}

There is a partition of the vertex set of $\graph(X)$ into $r$ cliques if and only if there is some ${\mathcal M} \in F(X)$ consisting of $r$ label sets. 
The clique covering number of graph  $\graph(X)$ can be obtained by
$\theta(\val(X)):=\min_{{\mathcal M} \in F(X)} |{\mathcal M}|$.

\begin{theorem} \label{nlcccn}
The clique covering number of a graph of
bounded clique-width can be computed in polynomial time.
\end{theorem}

\subsection{Computing  $\alpha$, $\omega$, $\chi$, and $\theta$ on co-graphs \label{sec-co}}

We next show that for graphs of clique-width at most 2 and graphs of NLC-width 1, i.e. for co-graphs (complement reducible graphs), our shown algorithms  can be simplified. A {\em co-graph} is either 
\begin{itemize}
\item
a  single vertex (denoted by $\bullet$),
\item
the disjoint union of two co-graphs $G_1,G_2$ (denoted by $G_1 \cup G_2$), or 
\item
the join of two co-graphs $G_1,G_2$, which connects every vertex of $G_1$ with every vertex of $G_2$ (denoted by $G_1 \times G_2$).
\end{itemize}
Obviously for every co-graph we can define a tree structure, denoted as {\em co-tree} in \cite{CPS85}. The leaves of the co-tree represent the vertices of the graph and the inner nodes of the co-tree  correspond to the operations applied on the
subexpressions defined by the two subtrees.  Given some co-graph $G$ we can construct a corresponding co-tree $T_G$ in linear time by the results shown in \cite{CPS85}. Using the tree structure $T_G$,
based on the results of Corneil et al. \cite{CLS81}, we next give simple linear time
algorithms for computing $\alpha$, $\omega$, $\chi$, and $\theta$ on co-graphs.


\subsubsection{Independence number \label{ssis}}

For every co-graph $G$ its independence number $\alpha(G)$ can recursively be computed as follows.

\begin{enumerate}
\item If $|V_G|=1$, then $\alpha(G):=1$.

\item If $G=G_1\cup G_2$, then $\alpha(G):=\alpha(G_1) + \alpha(G_2)$.

\item If $G=G_1\times G_2$, then $\alpha(G):=\max\{\alpha(G_1),\alpha(G_2)\}$.
\end{enumerate}

\begin{theorem} \label{co-in}
For every co-graph its independence number can be computed in linear time.
\end{theorem}

\subsubsection{Clique number\label{ssmc}}

For every co-graph $G$ its clique number $\omega(G)$ can recursively be computed as follows.

\begin{enumerate}
\item If $|V_G|=1$, then $\omega(G):=1$.

\item If $G=G_1\cup G_2$, then $\omega(G):=\max\{\omega(G_1),\omega(G_2)\}$.

\item If $G=G_1\times G_2$, then $\omega(G):=\omega(G_1) + \omega(G_2)$.
\end{enumerate}

\begin{theorem} \label{co-cn}
For every co-graph its clique number can be computed in linear time.
\end{theorem}

\subsubsection{Chromatic number}

Since co-graphs are perfect, we know that  for every induced subgraph $H$ of some
co-graph $G$ it holds $\chi(H)=\omega(H)$, and thus we can compute its chromatic number $\chi(G)$ by the same algorithm as shown for $\omega(G)$ above.

\begin{theorem}\label{co-chrn}
For every co-graph its chromatic number can be computed in linear time.
\end{theorem}

\subsubsection{Clique covering number}

Again, since co-graphs are perfect, we know that  for every induced subgraph $H$ of some
co-graph $G$ it holds $ \theta(H)=\alpha(H)$, and thus we can compute clique covering number $\theta(G)$ by the same algorithm as shown for $\alpha(G)$ above.

\begin{theorem}\label{co-clcovn}
For every co-graph its clique covering number can be computed in linear time.
\end{theorem}

\subsection{Complement problems\label{cw-cp}}

If ${\mathcal L} \subseteq \CW_k$ or  ${\mathcal L} \subseteq \NLC_k$, then for the corresponding set of complement graphs it holds
$\overline{\mathcal L}\subseteq  \CW_{2k}$ \cite{CO00} or $\overline{\mathcal L}\subseteq  \NLC_{k}$ \cite{Wan94}, respectively.
This implies that for every graph problem $\Pi$ solvable in polynomial time on clique-width bounded graphs, the corresponding complement problem $\overline{\Pi}$ is also solvable in polynomial time on clique-width bounded graphs.

For example, in order to solve the clique problem on clique-width bounded graphs one can use the data structure given in Section \ref{cwmaxcl}.
Alternatively, form a theoretically point of view, one could transform a given clique-width $k$-expression $X$ for
some given graph $G$ into a clique-width $2k$-expression $X'$ for its complement graph $\overline{G}$, and apply the algorithm for the independent set problem shown in Section \ref{cwis} on $X'$ in order to obtain the value of $\alpha(\overline{G})=\omega(G)$.
The same holds true for the solution of the partition into cliques problem on clique-width bounded graphs given in Section \ref{cwccn}. We can apply 
the algorithm for the partition into independent sets problem in Section \ref{ch534} on an expression for the complement graph in order to obtain the value of $\chi(\overline{G})=\theta(G)$.

From a practical point of view, one should prefer the solutions using a $k$-expression instead of those using
a $2k$-expression, since the clique-width of the input graph occurs as an exponent in the running time of our fpt algorithms.

\subsection{Clique-width and monadic second order logic}

On graph classes of bounded clique-width, all graph properties and optimization problems which are expressible in monadic second order logic with quantifications over vertices and vertex sets ($\MSOA$-logic) are decidable in linear 
time if a clique-width expression for the graph is given as an input \cite{CMR00}. 
This also implies the existence of linear time algorithms for computing the independence number and  clique number on graphs of bounded clique-width if a clique-width expression for the graph is given as an input. Note that the problems partition into independent sets and partition into cliques are not expressible in $\MSOA$-logic.

\section{Conclusions\label{ch-concl}}

Let us briefly discuss two further well known graph parameters which can be computed in polynomial 
time on graphs of bounded tree-width and graphs of bounded clique-width. 

A {\em dominating set} for some graph $G$ is a subset $S\subseteq V_G$, such that every vertex of $V_G-S$ is 
adjacent to at least one vertex from $S$.  The minimum value $s$ such that $G$ has a dominating set $S\subseteq V_G$ of size $s$ is denoted as 
the {\em dominating number} of graph $G$, denoted by $\gamma(G)$. 

In \cite{AP89} it is shown that the  dominating number of a graph of
bounded tree-width can be computed in linear time. Further in  \cite{Rao07a} it is shown that the dominating number of a graph of bounded clique-width can be computed in polynomial time.

\bigskip
A {\em vertex cover} for some graph $G$ is a subset $S\subseteq V_G$, such that every edge of $G$ has at least one endpoint in $S$. The minimum value $s$ such that $G$ has a vertex cover $S\subseteq V_G$ of size $s$ is denoted as 
the {\em vertex cover number} of graph $G$, denoted by $\tau(G)$. 

Gallai has shown in \cite{Gal59} the following relation between the size of a minimal vertex cover and maximum independent set of some graph $G$
\begin{equation} \label{gallai}
\tau(G)+\alpha(G)=|V_G|,
\end{equation}
which implies by Theorem \ref{twmis} that the  vertex cover number  of a graph of
bounded tree-width can be computed in linear time. For the same reason by  Theorem \ref{nlcmis} the  vertex cover number  of a graph of
bounded clique-width can be computed in linear time, if the
graph is given by some clique-width $k$-expression.

\bigskip
In this paper we have compared and illustrated how to use the tree structure of graphs of bounded tree-width and graphs of bounded clique-width to  give two general dynamic programming schemes to solve problems along a tree decomposition
and along a clique-width expression. Let us finally emphasize that
both approaches are useful. On the one hand, clique-width allows to define larger classes
of graphs of bounded width than tree-width. On the other hand, 
there are graph problems which remain NP-complete on graphs of bounded clique-width, but which are fixed-parameter tractable on graphs of bounded tree-width, such as the vertex disjoint path problem which is
discussed in \cite{GW06}. Further the tool of monadic second order logic allows
to define provable larger classes of problems which are solvable on tree-width bounded graph classes than on clique-width bounded graph classes, see  \cite{CMR00}.




\bibliographystyle{alpha}
\bibliography{/home/gurski/bib.bib}

\end{document}